\RequirePackage{fix-cm}
\RequirePackage{amsmath}
\documentclass[twocolumn,epjc3]{svjour3}  
\smartqed  
\RequirePackage{graphicx}
\RequirePackage{mathptmx}      
 
\usepackage{color}
\usepackage{mathrsfs}
\usepackage{amssymb, bm}

\newcommand{\be}{\begin{equation}}
\newcommand{\ee}{\end{equation}}

\begin{document}

\title{Revisiting geodesic observers in cosmology  
}


\author{Genevi\`eve Vachon\thanksref{e1,addr1}
	   \and
Robert Vanderwee  \thanksref{e2,addr1} 
	   \and
Valerio Faraoni  \thanksref{e3,addr1} 
}

\thankstext{e1}{e-mail: gvachon18@ubishops.ca}
\thankstext{e2}{e-mail: rvanderwee20@ubishops.ca}
\thankstext{e3}{e-mail: vfaraoni@ubishops.ca}


\institute{Department of Physics \& Astronomy, Bishop's University, 
2600 College Street, Sherbrooke, Qu\'ebec, Canada J1M~1Z7 \label{addr1}
}

\date{Received: date / Accepted: date}

\maketitle

\begin{abstract} 

Geodesic observers in cosmology are revisited. The coordinates based on 
freely falling observers introduced bu Gautreau in de Sitter and 
Einstein-de Sitter spaces (and, previously, by Gautreau \& Hoffmann in 
Schwarzschild space) are extended to general FLRW universes. We identify 
situations in which the relation between geodesic and comoving coordinates 
can be expressed explicitly in terms of elementary functions. In general, 
geodesic coordinates in cosmology turn out to be rather cumbersome and 
limited to the region below the apparent horizon.


\end{abstract}

\section{Introduction}
\label{sec:1}
\setcounter{equation}{0}

Geodesic observers in radial free fall, and the associated coordinates, 
were introduced in Schwarzschild spacetime long ago by Ronald Gautreau and 
Banesh Hoffmann \cite{Gautreau:1978zz} (see also 
\cite{FtaclasCohen,Gautreau:1984pny,Gautreau95,Gautreau00,Francis:2003rj,Finch:2012vli,Mitra:2013rma,Posada:2014xda} 
and  
Refs.~\cite{MacLaurin:2018aze,MacLaurin:2019gpc,MacLaurin:2019scl,Lemos:2020qxk,Grib:2020kzh} 
for recent interest).  Gautreau used them also in 
Friedmann-Lema\^itre-Robertson-Walker (FLRW)  cosmology 
\cite{Gautreau83,Gautreau84}. Gautreau's \cite{Gautreau84} key idea was to 
use freely falling observers to describe spatially homogeneous and 
isotropic cosmology, therefore the Gautreau-Hoffmann coordinates in 
Schwarzschild \cite{Gautreau:1978zz} and their analogue in FLRW 
\cite{Gautreau83,Gautreau84} spacetimes should properly be called 
``geodesic coordinates''. Gautreau's motivation for using these 
coordinates in cosmology in his 1983 \cite{Gautreau83} and 1984 
\cite{Gautreau84} papers remains rather obscure, since it is far more 
natural to describe cosmology from the point of view of comoving observers 
(those that see the cosmic microwave background spatially homogeneous and 
isotropic around them, apart from the tiny temperature perturbations 
$\delta T/T_0 \simeq 5 \cdot 10^{-5}$ discovered by the {\em COBE} satellite 
in 1992). 
However, today there is a large literature on the mechanics and thermodynamics 
of apparent horizons which often require alternative 
coordinates.\footnote{See, for 
example, Refs.~\cite{Parikh:1999mf,Hong:2020dow} for the use of 
Painlev\'e-Gullstrand coordinates to describe the thermodynamics of the 
Schwarzschild horizon and Ref.~\cite{Parikh:2002qh} for the de 
Sitter horizon.} Cosmological horizons are increasingly studied as almost 
trivial 
examples of apparent horizons to test properties of the analogous 
(but more complicated) apparent horizons of dynamical black holes. Moreover,  
cosmology has expanded significantly with 1)~ the inflationary paradigm of the 
early universe; 2)~the 
discovery of cosmic microwave background temperature fluctuations  in 
1992, and 3)~the 1998 discovery, made with type Ia supernovae, of the present 
acceleration of the cosmic expansion. This significant growth of cosmology and 
of horizon mechanics and thermodynamics motivates the exploration of subjects 
that were marginal in the past, in particular contemplating alternative 
coordinate systems in cosmology is more motivated today than it was in the 
1980s.

Gautreau \cite{Gautreau84} restricted himself to spatially flat  universes, 
then 
further restricted to Einstein-de Sitter universes in which the fluid is a 
dust and the comoving observers are geodesic \cite{Gautreau84}, or to an 
empty and locally static de Sitter universe \cite{Gautreau83} with 
positive cosmological constant $\Lambda$. Then, he further restricted himself 
to the discussion of geodesic observers starting their radial free fall from 
the origin $r=0$. We would like to go beyond all these limitations.

The approach of Refs.~\cite{Gautreau83} and~\cite{Gautreau84} is rather 
indirect: Gautreau first writes the FLRW line element as a generic 
spherically symmetric one using the areal radius as the radial coordinate, 
and then solves the Einstein equations. Only later, spatial homogeneity and 
isotropy are imposed. There is no need to do this as the FLRW geometry 
describing spatial homogeneity and isotropy is well known \cite{Wald}. 
Probably due to the lack of a widespread geometric view at the time of 
writing,\footnote{For example, statements such as ``the time coordinate is 
not suitable for formulating the cosmological principle'' and ''is tied to 
one particular galaxy'' would lead a reader to suspect that spatial 
homogeneity and isotropy are coordinate-dependent properties, while 
spacetime symmetries are instead intrinsic and, indeed, comoving 
coordinates (including comoving time) are adapted to these 
symmetries.} Gautreau's papers \cite{Gautreau83,Gautreau84} are rather 
obscure on several points 
that can use a transparent geometric clarification or reformulation. In 
several other points the reasoning is vague or borderline incorrect (for 
example, comoving 
observers are confused with geodesic ones, although this no longer matters 
when Gautreau specializes to a dust fluid, but becomes crucial when 
attempting to move beyond this limitation). Certain reasonings are 
ultimately correct, but this can only be established {\em a posteriori}. 
As a result, the average reader would remain suspicious about the 
derivation 
of geodesic coordinates in \cite{Gautreau83,Gautreau84} and would avoid using 
them. 

Here we revisit critically the Gautreau construction of geodesic coordinates 
and we 
attempt to give a more direct and transparent treatment, while removing 
the heavy restrictions of Refs.~\cite{Gautreau84} and 
\cite{Gautreau83}. We begin by using the FLRW geometry in comoving 
coordinates from the outset, then transforming to Gautreau-Hoffmann-like  
coordinates employing the areal (or ``curvature'', or ``Schwarzschild-like'') 
radius as the radial coordinate and the proper time of radial geodesic 
observers as the time coordinate. We elucidate several points not 
addressed in Refs.~\cite{Gautreau83,Gautreau84}. As will be clear in the 
following sections, connecting geodesic coordinates with the more natural 
comoving coordinates cannot always be done explicitly, in particular for 
spatially curved FLRW universes. We highlight situations in which the 
relation between geodesic and comoving time can be calculated explicitly in 
terms of elementary functions, and we provide explicit examples of 
physical interest. It turns 
out that geodesic coordinates in FLRW cosmology are rather cumbersome and 
only cover the region of FLRW space below the apparent horizon. 
Indeed, the discussion of radial geodesic observers quickly becomes very 
involved and, to keep it manageable, we will restrict ourselves to 
observers initially comoving with the cosmic fluid. 
Likewise, we only consider FLRW universes sourced by a single perfect 
fluid in the context of Einstein's theory of gravity.  

We follow the notation of Ref.~\cite{Wald}: the metric signature is 
${-}{+}{+}{+}$ and we use units in which Newton's constant $G$ and the 
speed of light $c$ are unity.

\section{Geodesic and quasi-geodesic observers in FLRW universes} 
\label{sec:2}

Here we introduce geodesic coordinates in FLRW spacetime, which are analogous 
to the Gautreau-Hoffmann coordinates of Schwarzschild spacetime. 
These new geodesic coordinates in the FLRW universe are associated with 
observers in radial free fall and starting from rest with respect to the 
cosmic fluid.

Let us begin by illustrating the difference between geodesic and 
quasi-geodesic ({\em i.e.}, comoving) observers in FLRW space. In general, 
the worldlines of freely falling observers are (timelike) spacetime 
geodesics. When 
the spacetime is sourced by a single perfect fluid, the worldlines of the 
fluid parcels differ from those of geodesic observers, unless the fluid is 
a dust \cite{Wald} because of the pressure gradient $\nabla_a P$ acting on 
them. Specializing to FLRW spacetimes, quasi-geodesics are defined as 
worldlines which are identical in appearance to a geodesic, yet they 
differ by the fact that their proper time is a non-affine parameter 
\cite{Faraoni:2020ehi}. A quasi-geodesic is the worldline of a particle 
subject to a pressure gradient parallel to the particle four-velocity (it 
is clear that, because of spatial isotropy, the four-gradient $\nabla_c P$ 
of the pressure $P(t)$ points in the time direction of comoving observers 
in a FLRW universe). A quasi-geodesic observer perceives a 3-space which 
is Lorentz-boosted relative to the proper 3-space of a geodesic observer  
\cite{Faraoni:2020ehi}. It was previously shown in 
\cite{Faraoni:2020ehi} that the motion of a fluid particle in a FLRW 
universe is a radial timelike quasi-geodesic.

The FLRW line element in comoving polar coordinates $\left( t, r, 
\vartheta, \varphi \right)$ is \cite{Wald}
\be 
ds^2=-dt^2 +a^2(t) \left( \frac{dr^2}{1-kr^2} +r^2 d\Omega_{(2)}^2 \right) 
\,,\label{FLRWcomoving}
\ee
where $a(t)$ is the scale factor describing the expansion history of the 
universe, $k$ is the curvature index normalized to $0, \pm 1$, and $ 
d\Omega_{(2)}^2\equiv d\vartheta^2 +\sin^2\vartheta \, d\varphi^2$ is the 
line element on the unit 2-sphere. We then introduce the areal radius 
$R(t,r) \equiv a(t) r$, which is analogous to the Schwarzschild radius and 
is the radial coordinate in the Gautreau coordinate system 
\cite{Gautreau83,Gautreau84}. In 
principle, in a FLRW universe one could also use the proper radius defined 
by
\begin{equation}
 R_p = a(t)\int{\frac{dr}{\sqrt{1-kr^2}}} \equiv a(t)\chi \,,
\end{equation}
where $\chi$ is the hyperspherical radius often used in cosmology and   
\begin{equation}
f(\chi) = r \equiv 
    \begin{cases}
        \sin \chi  \quad &\text{if   } \, k=+1 \,,\\
        \chi \quad &\text{if   } \, k=0 \,,\\
        \sinh \chi  \quad &\text{if   } \, k=-1 \,,
    \end{cases}
\end{equation}
which turns the FLRW line element~(\ref{FLRWcomoving}) into 
\begin{equation}
ds^2 = -dt^2+a^2(t) \left( d\chi^2+f^2(\chi)d\Omega_{(2)}^2 \right) \,.
\end{equation}
$R_p$ is a ``volume radius'' rather than an areal radius and coincides 
with $R$ only for spatially flat ($k=0$) universes. Gautreau \& Hoffmann 
\cite{Gautreau:1978zz} used the 
areal radius $R$ instead of the proper radius $\int{\frac{dR}{\sqrt{1- 
2M/R}}}$ in the Schwarzschild geometry 
\be
ds^2 = -\left(1- \frac{2m}{R} \right) dt^2 + \frac{dR^2}{1- 2m/R} 
+ R^2 \Omega_{(2)}^2 \,,
\ee
hence the analogue of their coordinates in FLRW space should use the areal 
radius  $R$ as 
well. Defining the proper time of geodesic observers and linking it with 
the time coordinate $t$ of comoving observers is considerably more 
complicated than introducing the areal radius $R$.

\subsection{Timelike radial geodesics in FLRW}

Let us begin by characterizing the radial timelike geodesics of FLRW 
spacetime. The non-vanishing Christoffel symbols of the FLRW geometry in 
comoving coordinates $\left(t, r, \vartheta, \varphi \right)$ are 
\begin{eqnarray}
\Gamma_{rr}^t &=& \frac{a\dot{a}}{1-kr^2} \,,\\
&&\nonumber\\
\Gamma_{\vartheta \vartheta}^t &=& a\dot{a}r^2 \,,\\
&&\nonumber\\
\Gamma_{\varphi\varphi}^t &=& a \dot{a} r^2 \sin^2 \vartheta \,,\\
&&\nonumber\\
\Gamma_{tr}^r &=& \Gamma_{rt}^r= \Gamma_{t\vartheta}^\vartheta =
\Gamma_{\vartheta t}^\vartheta = 
 \Gamma_{t\varphi}^\varphi = \Gamma_{\varphi t}^\varphi 
= \frac{\dot{a}}{a} \,,\\
&&\nonumber\\
\Gamma_{\vartheta \vartheta}^r &=&   -r\left( 1-kr^2 \right) \,,\\
&&\nonumber\\
\Gamma_{\varphi \varphi}^r &=& -r\left(m 1-kr^2 \right)\sin^2 \vartheta  \,,\\
&&\nonumber\\
\Gamma_{\varphi \varphi}^\vartheta &=&  -\sin \vartheta \cos \vartheta \,,\\
&&\nonumber\\
\Gamma_{\vartheta \varphi}^\varphi &=& \Gamma_{\varphi \vartheta}^\varphi 
= \cot \vartheta \,,\\
&&\nonumber\\
\Gamma_{r \vartheta}^\vartheta &=&\Gamma_{\vartheta r}^\vartheta =
\Gamma_{r\varphi}^\varphi =  \Gamma_{\varphi r}^\varphi = \frac{1}{r} \,, 
\end{eqnarray}
where an overdot denotes differentiation with respect to the comoving 
time $t$. A radial timelike geodesic with proper time $\tau$ and 
four-velocity components 
\be
u^{\mu} = \frac{dx^{\mu}}{d\tau} =\left( u^t, u^r, 0,0 
\right)
\ee
satisfies the geodesic equations
\begin{eqnarray}
&&    \frac{du^r}{d\tau} + 2\Gamma_{tr}^r \, u^t u^r = 0 \,,\\
&&\nonumber\\
&& \frac{du^r}{d\tau} + 2\, \frac{\dot{a}}{a} \, u^tu^r = 0 \,.
\end{eqnarray}
Dividing by $u^t = dt/d\tau $, one obtains 
\begin{equation}
    \frac{d}{dt}\Bigg[ \ln(u^r)+2\ln\bigg(\frac{a}{a_0}\bigg)\Bigg] = 0 
\,,
\end{equation}
which integrates to 
\be
    u^r = u_{(0)}^r \, \frac{a_0^2}{a^2} \label{solution:u^r1} \,,
\ee
where $u_{(0)}^r \equiv u^r(t_0)$ is the initial condition at the comoving 
time $t_0$ and $a_0 \equiv a(t_0)$. The normalization of the four-velocity 
$g_{ab} u^a u^b =-1$ gives
\be
  -\left(u^t\right)^2 + \frac{a^2}{1-kr^2}\left(u^r \right)^2 = -1
\ee
and Eq.~(\ref{solution:u^r1}) then yields
\begin{align}
u^t = \sqrt{1+\frac{(u_{(0)}^r)^2a_0^4}{a^2(1-kr^2)}} \,,
\label{solution:u^t1}
\end{align}
where the positive sign of the square root is chosen in order for $u^a$ to 
be future-oriented.  
If the geodesic particle is initially at rest in comoving 
coordinates ({\em i.e.}, initially comoving with the cosmic fluid) at 
 time $t_0$ and position\footnote{Note that the initial radius of the geodesic 
observer is not restricted to vanish, as in \cite{Gautreau83,Gautreau84}.}  $ 
x_{(0)}^\mu = \left( t_0, r_0,\vartheta_0,\varphi_0 \right)$, then the 
components of its four-velocity are
\begin{equation}
    u_{(0)}^\mu = \left( 1,0,0,0 \right) \,,
\end{equation}
that is, the four-velocity coincides with that of a radial timelike 
geodesic. In other words, {\em if the freely falling particle is initially 
comoving with the cosmic fluid, it remains comoving at all times} 
\cite{Faraoni:2020ehi}. This point was missed in 
Refs.~\cite{Gautreau83,Gautreau84}. The time component of the geodesic 
equation then becomes
\be
\frac{du^t}{d\tau} +\frac{a\dot{a}}{1-kr^2} \left( u_{(0)}^r \right)^2 \, 
\frac{a_0^4}{a^4} =0 \,,
\ee
which integrates to 
\be
u^t=\frac{dt}{d\tau} = \alpha \, t + \beta 
\ee
(with $\alpha$ and $\beta$ integration constants) and
\be
t(\tau) = \frac{\alpha\, \tau^2}{2} + \beta \, \tau +\gamma \,,
\ee
where $\gamma$ is another integration constant. For the FLRW cosmic fluid, 
$t$ is the proper time of the fluid particles, while $\tau$ is the proper 
time of massive test particles: the two do not coincide unless the fluid 
is a dust.

\subsection{Pseudo-Painlev\'e-Gullstrand coordinates}

By switching from comoving radius $r$ to the areal radius $R(t,r) \equiv 
a(t)r$, and using the relation between differentials $dr=\left( dR 
-HRdt\right)/a$, the FLRW line element~(\ref{FLRWcomoving})  assumes 
the non-diagonal form (dubbed ``pseudo-Painlev\'e-Gullstrand'' 
form\footnote{This line element resembles 
the Painlev\'e-Gullstrand line element for the Schwarzschild geometry but,  
unless $k=0$, it lacks the defining feature of Painlev\'e-Gullstrand 
coordinates that the constant time slices are flat 
\cite{Painleve,Gullstrand,Martel:2000rn}.} \cite{Faraoni:2015ula})
\begin{eqnarray}
ds^2 &=& -\left(1-\frac{H^2R^2}{1-kR^2/a^2} \right)dt^2  - 
\frac{2HR}{1-kR^2/a^2} \, dtdR \nonumber\\
&&\nonumber\\
&\, &  + \, \frac{dR^2}{1-kR^2/a^2} + R^2 d\Omega_{(2)}^2 \,,\label{pseudoPG}
\end{eqnarray}
where $H \equiv \dot{a}/a$ is the (comoving time) Hubble function. 
In these coordinates, the four-velocity normalization reads
\begin{eqnarray}
&& \left(-1 + \frac{H^2R^2}{1-kR^2/a^2}\right)(u^t)^2
-\frac{2HR}{1-kR^2/a^2}u^tu^R  \nonumber\\
&&\nonumber\\
&& +\frac{(u^R)^2}{1-kR^2/a^2} = -1  \label{relation:4vel1}
\end{eqnarray}
and can be rewritten in the form 
\begin{equation} 
-\left(u^t \right)^2 + \frac{1}{1 - k R^2 / a^2} 
\left(u^R - HR u^t \right)^2 = 0   
\label{relation:4vel2} 
\end{equation}
that will be useful later. Eq.~(\ref{relation:4vel1}) is solved for $u^R$, 
yielding the quadratic equation  
\begin{eqnarray}
&&    (u^R)^2-2HR \, u^tu^R-\left(1-H^2R^2 
-\frac{kR^2}{a^2}\right)(u^t)^2+1-\frac{kR^2}{a^2}  \nonumber\\
&& =0 
\end{eqnarray}
with roots
\begin{eqnarray}
u^R & = & HR \, u^t \nonumber\\
&&\nonumber\\
& \pm & \sqrt{H^2R^2(u^t)^2 +
\left( 1-H^2R^2-\frac{kR^2}{a^2}\right)(u^t)^2-1+\frac{kR^2}{a^2}} 
\,.\nonumber\\
&&
\end{eqnarray}
The argument of the  square root can be rewritten as 
\be
 \left(1-\frac{kR^2}{a^2}\right)[(u^t)^2-1] \,,
\ee 
so that
\begin{align}
    u^R = HR \, u^t \pm \sqrt{1-\frac{kR^2}{a^2}}\sqrt{(u^t)^2-1} \,.
\end{align}
We can now relate the components of the four-velocity in 
pseudo-Painlev\'e-Gullstrand coordinates to those in comoving coordinates. 
Since
\begin{equation}
u^R \equiv \frac{dR}{d\tau} =  
\dot{a} \, \frac{dt}{d\tau} \, r + a\, \frac{dr}{d\tau} = 
\frac{\dot{a}}{a} \, Ru^t + a u^r 
\end{equation}
and
\be
u^r = \frac{u^R}{a}-\frac{HR}{a}\, u^t \,, \quad\quad  u^R = H R \, 
u^t + a \, u^r \,, \label{relation:u^R}
\ee
applying Eqs.~(\ref{solution:u^r1}) and~(\ref{solution:u^t1}) to the 
second of Eqs.~(\ref{relation:u^R}) gives 
\begin{align}
 u^R = HR\sqrt{1+\frac{(u_{(0)}^r)^2 \, a_0^4}{ 
a^2(1-kr^2)}} \, \pm  u_{(0)}^r \, \frac{a_0^2}{a} \,. \label{solution:u^R1}
\end{align}
We now impose the special initial condition\footnote{Gautreau imposes the 
special 
initial position $R_0=0$ invokingthe cosmological principle--the meaning of 
this statement is unclear. We do not impose this unnecessary restriction and 
the geodesic clock can be dropped from any initial position below the 
apparent horizon.} 
\begin{eqnarray} 
R(t_0) &=& R_0 \,,\label{IC1}\\ 
&&\nonumber\\ u_{(0)}^R &=& 0 \,,\label{IC2}
\end{eqnarray} 
at $t=t_0$  (or $ \tau = \tau_0$). Physically, this means that the geodesic 
clock is released from rest at $R_0$, where ``at rest'' means $ dR/d\tau 
\equiv u^R =0 $. With  
these initial conditions, the normalization~(\ref{relation:4vel1}) gives
\begin{equation}
\left(1-\frac{H_0^2R_0^2}{1-k 
R_0^2/a_0^2}\right)(u_{(0)}^t)^2 = 1
\end{equation}
and the initial time component 
\begin{align}
u_{(0)}^t = \sqrt{\frac{1-k R_0^2/a_0^2}{1-H_0^2R_0^2-k 
R_0^2/a_0^2}} \,.\label{solution:u^t2}
\end{align}
Substituting this expression into the first of Eq.~(\ref{relation:u^R}) 
yields
\begin{equation}
    u_{(0)}^r = -\frac{H_0 \, R_0 \, u_{(0)}^t}{a_0} 
\end{equation}
and, finally, 
\begin{align}
u_{(0)}^r = -\frac{H_0R_0}{a_0} \, \sqrt{\frac{1-k\, R_0^2/a_0^2}{ 
1-H_0^2R_0^2-k R_0^2/a_0^2}} \,. \label{solution:u^r2}
\end{align}
Eq.~(\ref{solution:u^r2}) agrees with what one obtains by setting 
$u^R_0 = 0$ in Eqs.~(\ref{solution:u^R1}) and (\ref{relation:4vel2}).
Using the normalisation of the four-velocity in comoving coordinates and 
Eq.~(\ref{solution:u^r2}), one obtains
\begin{equation}
   u^t = \sqrt{\frac{a^2_0 \, H^2_0 \, R^2_0}{ a^2 \left(1 - k 
R^2/a^2 \right)}  \frac{1 - k R_0^2/a_0^2}{1 - k 
R_0^2/a_0^2 - H^2_0 R^2_0 } + 1}  
\label{solution:u^t3}
\end{equation}
along the radial timelike geodesics with the special initial 
condition~(\ref{IC1}), (\ref{IC2}). As a check, Eq.~(\ref{solution:u^t3}) 
agrees with Eq.~(\ref{solution:u^t2}) at the spacetime point $\left( t_0, 
R_0,  \vartheta_0, \varphi_0 \right)$.

The use of Eqs.~(\ref{solution:u^R1}) and~(\ref{solution:u^r2}) then leads to
\begin{align}
u^R =&  HR\, \sqrt{1 + \frac{a_0^4 \left(u_{(0)}^r \right)^2}{a^2\left(1- 
- kr^2 \right)}} \, \pm u_{(0)}^r \, \frac{a_0^2}{a^2} \nonumber\\
&\nonumber\\
=& HR \,  \sqrt{1 + \frac{a_0^4}{a^2\left( - kr^2 \right)}\, \frac{H_0^2 
\, R_0^2}{a_0^2} \,  \frac{1 - k R_0^2 / a_0^2}{1 - k R_0^2 
/ a_0^2 - H_0^2 R_0^2}} \nonumber\\
&\nonumber\\
& \pm  u_{(0)}^r \, \frac{a_0^2}{a^2} \nonumber\\
&\nonumber\\
 =& HR \, \sqrt{1 + \frac{a_0^2 H_0^2 R_0^2}{a^2\left(1 - kr^2\right)} 
\, \frac{1 - k R_0^2 / a_0^2}{1 - k R_0^2 / a_0^2 - 
H_0^2 R_0^2}} \nonumber\\
&\nonumber\\
&  \mp \frac{H_0 \, R_0 \, a_0}{a} \, \sqrt{\frac{1 - k R_0^2 
/ a_0^2}{1 - k R_0^2 / a_0^2 - H_0^2 R_0^2}} \,, 
\end{align}
which (as a check) satisfies $u_{(0)}^R = 0$ at $R_0$. For an observer 
initially at rest 
({\em i.e.}, $u_{(0)}^R = 0$), the relation~(\ref{relation:u^R}) 
suggests that 
\be
u_{(0)}^r = -\frac{H_0 \, R_0 \, u_{(0)}^t }{a_0} < 0\, :
\ee 
of course, if this geodesic observer is at rest in the Gautreau-Hoffmann 
sense, it is left behind by the comoving observers and its radial 
velocity according to the comoving observers is negative.

In the following we need the components of the four-velocity covector  
\begin{align}
    u_t =& \ g_{t \alpha} u^\alpha \nonumber \\
    \nonumber\\
    =& \ -\left(1-\frac{H^2R^2}{1-kR^2/a^2}\right)  u^t - 
\frac{HR}{1-kR^2/a^2} u^R  \nonumber \\
    \nonumber\\
    =& \ \frac{a_0 \, H_0 \, R_0 \, H\, R}{a \left(1 - k R^2 /a^2 \right)} 
\sqrt{\frac{1 - k R_0^2/a_ 0^2}{1 - k R_0^2/a_0^2 - 
H^2_0 R^2_0 }} \nonumber\\
&\nonumber\\
    & \ -\sqrt{1 + \frac{a_0^2 \, H_0^2 \, R_0^2}{a^2 \left(1 -  k 
R^2 / a^2 \right)}\,  \frac{1 - k R_0^2 / a_0^2}{1 - k R_0^2 / 
a_0^2 - H_0^2 R_0^2}} 
\end{align}
and
\begin{align}
    u_R =& \ g_{R \alpha} u^\alpha, \nonumber\\
    =& \mp \frac{HR}{1-kR^2/a^2} \, u^t + \frac{1}{1-kR^2/a^2} \, u^R 
\nonumber\\
    =&  \mp \frac{a_0 \, H_0 \, R_0 }{a \left(1 - k R^2 /a^2 \right)} 
\sqrt{\frac{1 - k R_0^2/a_0^2}{1 - k R_0^2/a_0^2 - 
H^2_0 R^2_0 }} \,.
\end{align}

\subsection{Geodesic coordinates}

The Gautreau-Hoffmann-like geodesic coordinates are\\ $\left( \tau, R, 
\vartheta, \varphi \right)$, where $\tau$ is the proper time of clocks freely 
falling from rest ({\em i.e.}, $u_{(0)}^R = 0$ initially). The relation 
between $\tau$ and the comoving time $t$ is given by 
$ u^t \equiv dt/d\tau $ and $ d\tau = dt/u^t $. In finite terms, 
\begin{eqnarray}
\tau &=& = \int{\frac{dt}{u^t}}  \label{relation:ttau}\\
&&\nonumber\\
&=& \int{dt \left[ \frac{a^2_0 \, H^2_0 \, R^2_0 }{a^2 \left(1 
- k R^2/a^2 \right)} \right.}\nonumber\\
&&\nonumber\\
& \, & \left. \times \frac{1 - k R_0^2/a_0^2}{1 - k 
R_0^2/a_0^2 - H^2_0 \, R^2_0 } + 1 \right]^{-1/2} \nonumber\\
&&\nonumber\\
&=& \int{dt \sqrt{\left(a^2 - kR^2 \right) \left(1 - \frac{k 
R_0^2}{a_0^2} - H^2_0 \, R^2_0 \right)}} \nonumber\\
&&\nonumber\\
 & \, & \times \left[\left(a^2 - kR^2 \right) \left(1 - \frac{k 
R_0^2}{ a_0^2} - H^2_0 \, R^2_0 \right) \right.\nonumber\\
&&\nonumber\\
& \, & \left. + a^2_0 \, H^2_0 \, R^2_0 \left(1 - \frac{k 
R_0^2}{a_0^2} \right) \right]^{-1/2} \,.
\end{eqnarray}
Using the notation 
\begin{eqnarray}
\alpha_0 &\equiv & 1 - \frac{k R_0^2}{a_0^2} - H^2_0 R^2_0 \,,\label{alpha0}\\
&&\nonumber\\
\beta_0 &\equiv & H^2_0 R^2_0 \left(a_0^2- k R^2_0 \right)\,, \label{beta0}
\end{eqnarray}
the $\tau$-coordinate is expressed by the integral 
\begin{equation}
\tau = \sqrt{\alpha_0} \, \int{dt \, \sqrt{\frac{a^2 - kR^2}{\left(a^2 - kR^2 
\right) 
\alpha_0 + \beta_0 }}} \,,
\end{equation}
where\footnote{Switching to conformal time does not help in computing this 
integral.} $a=a(t)$.

\section{Spatially flat FLRW universes}
\label{sec:3}

Motivated by modern cosmological observations, let us restrict to a 
spatially flat FLRW universe. For $k=0$, the Gautreau-Hoffmann-like geodesic  
time reduces to
\begin{equation}
\tau = \sqrt{\alpha_0} \, \int{dt \sqrt{\frac{a^2}{\alpha_0 \, a^2 + \beta_0 
}}} \,,
\end{equation}
and it is sometimes possible to express it in terms of elementary 
functions. Below, we discuss these integrability situations.

\subsection{Power-law scale factor}

Let us consider first a power-law scale factor, which always occurs for 
a spatially flat FLRW universe dominated by a single perfect fluid with 
constant barotropic equation of  state $P=w\rho$ \cite{Wald}, 
\begin{equation}
    a(t) = a_* t^p \,,
\end{equation}
where $a_*$ is a constant. In this case, it is 
\begin{equation}
\tau = \sqrt{\alpha_0 } \, a_* \int{dt \, t^p \left( \alpha_0 \, a_*^2 
t^{2p} + \beta_0 \right)^{-1/2}} \,.
\end{equation}
According to the Chebysev theorem of integration 
\cite{Chebysev,MarchisottoZakeri}, the integral  
\be
\int{dt \ t^p \left(A + B \, t^r \right)^q} \label{Cintegral}
\ee
where $A,B, p,q,, r$ are constants and $r\neq 0$, $p,q, r\in \mathbb{Q}$,   
is expressed in terms of a finite 
number of  elementary functions if and only if at least one of 
\be
\frac{p + 1}{r} \,, \quad q \,, \quad\quad \frac{p + 1}{r} + q
\ee
is an integer \cite{Chebysev,MarchisottoZakeri}.

An alternative approach consists of using a representation of the 
integral~(\ref{Cintegral}) in terms of a 
hypergeometric series and noting that the assumptions of the Chebysev 
theorem are equivalent to the condition for this series to reduce to a 
finite sum (this equivalent condition was noted several times in the context 
of two-fluid cosmologies, for which the Friedmann equation reduces to an 
integral of the same type 
\cite{Jacobs1968,McIntosh1972,McIntoshFoyster1972,Chen:2014fqa,Faraoni:2021opj}). 

In our case we can 
assume $p \in \mathbb{Q}$. In general, if the equation of state of the 
cosmic fluid has the barotropic form $P = 
w \rho$ with $w=$~const., then
\begin{equation}
a(t) = a_* \, t^{ \frac{2}{3(w+1) }} \,,
\end{equation}
and $w \in \mathbb{Q}$ implies that $p = 2/[3(w+1)] \in \mathbb{Q}$. 
Most values of the equation of state parameter $w$ used in the 
cosmological literature are indeed rational but, if 
this is not the case, one can always approximate $w \in \mathbb{R}$ with 
its rational approximation, still satisfying the cosmological 
observations to the required precision. We have then that $ p , r=2p, q=-1/2 
\in \mathbb{Q}$ and  
\begin{align}
  \frac{p + 1}{r} = \frac{p + 1}{2p}\,, && q = -\frac{1}{2} \notin 
\mathbb{Z}\,, 
&& \frac{p + 1}{r} + q = \frac{1}{2p} \,;
\end{align}
it is 
\be
\frac{p + 1}{r} = m \in \mathbb{Z}
\ee
if and only if $p = 1/(2m - 1)$, while  
\be
\frac{p + 1}{r} + q = m  \in \mathbb{Z}
\ee
if and only if $p  = 1/2 m$, so at least one of $(p + 1)/r, \ q, \ (p + 
1)/r + q \in \mathbb{Z}$ if $p = 1/n$, where $n =\pm 1, \pm2, \pm3 , 
\,...$  This list includes several well known equations of state in 
cosmology. Setting 
\be
p = \frac{2}{3(w+1)} \,, \quad\quad w_n = \frac{2n - 3}{3} \,, 
\ee
we have the equations of state listed in Table~\ref{tab:n}.
 
\begin{table}[h]
    \centering
    \caption{The equation of state parameter for a few values of $n$.}
    \begin{tabular}{|c||c c c c c|}
         \hline
         \textit{n} & $-1$ & $1$ & $2$ & $3$ & ... \ \\
         \hline
         \textit{w} & $-5/3$ & $-1/3$ & $\ 1/3$ & $\ \ 1 \ $ & ... \ \\
         \hline
    \end{tabular}
    \label{tab:n}
\end{table}

Let us discuss, as examples, two of these integrability cases, plus one not 
given by the Chebysev theorem.

\subsection{Radiation fluid}

The equation of state and power-law scale factor associated with a 
$k=0$ FLRW universe filled with radiation are $
    P = \rho/3$ and $a(t) = a_* \sqrt{t}$, which give
\begin{eqnarray}
\tau & = & \ \sqrt{\alpha_0} \, a_* \int{dt \ \sqrt{t} 
\left(\alpha_0 \, a_*^2 \, t + \beta_0 \right)^{-1/2}} \nonumber\\
&&\nonumber\\
&=& \ \frac{1}{\alpha_0 \, a_*^2} \left[\sqrt{\alpha_0 \, a_*^2 \, t + 
\beta_0 } \, \sqrt{\alpha_0 \, a_*^2 \, t} \right. \nonumber\\
&&\nonumber\\
& \, & \left.  - \beta_0 \sinh^{-1} \left(\sqrt{\frac{\alpha_0 \,   
a_*^2 \, t}{\beta_0 }} \right)\right] \,.
\end{eqnarray}

\subsection{Stiff fluid/free scalar field}

A universe filled with a stiff fluid with equation of state\footnote{It is 
well known that a stiff fluid is equivalent to a free scalar field.} $P = 
\rho$ has scale factor $ a(t) = a_* \, t^{1/3}$, yielding 
\begin{eqnarray}
    \tau &=&  \sqrt{\alpha_0 } \, a_* \int{dt \ t^{1/3} 
\left[ \alpha_0 \, a_*^2 \, t^{2/3} + \beta_0 \right]^{-1/2}} \\
&&\nonumber\\
&=& \frac{1}{\alpha_0^{3/2} \, a_*^3}\left(\alpha_0 \, a_*^2 \,
t^{2/3} + 2\beta_0 \right) \sqrt{ \alpha_0 \, a_*^2 \, t^{2/3} + 
\beta_0 } \,.
\end{eqnarray}

\subsection{de Sitter space}

The de Sitter universe with scale factor $ a(t) = a_* \, \mbox{e}^{Ht}$, $ 
H=$~const. is  another special case in which the Gautreau-Hoffmann-like  
geodesic time can 
be computed explicitly, giving 
\begin{eqnarray}
\tau &=& \sqrt{ \alpha_0 } \, a_* \int{dt \ \mbox{e}^{Ht} 
\left( \alpha_0 \, a_*^2 \, \mbox{e}^{2Ht} + \beta_0 \right)^{-1/2}} \\
&&\nonumber\\
&=& \frac{1}{H} \, \tanh^{-1}{ \left( \frac{ \mbox{e}^{Ht}}{ 
\sqrt{ \mbox{e}^{2Ht} + 
\frac{ \beta_0 }{ \alpha_0 \, a_*^2 } }} \right) } \,.
\end{eqnarray} 
The relation $\tau = \tau(t)$ can be inverted to find $t(\tau)$ and $a(\tau)$: 
from
\be
\frac{ \mbox{e}^{Ht}}{\sqrt{ \mbox{e}^{2Ht} + \beta_0/\left( \alpha_0  
\, a_*^2  \right)}} = \tanh{\left(H\tau \right)} 
\ee
one obtains 
\be
\mbox{e}^{2Ht} \left[ 1 - \tanh^2{\left( H\tau \right)} \right] =  
\frac{\beta_0}{\alpha_0 \, a_*^2} \, \tanh^2{ \left( H\tau \right)}
\ee
and then
\be
\mbox{e}^{Ht} = \sqrt{\frac{\beta_0 }{\alpha_0 \, a_*^2}} \, 
\sinh{\left( H\tau \right)} \,,
\ee
and taking the logarithm gives
\be
t =  \frac{1}{H} \, \ln{\left[\sqrt{\frac{\beta_0 }{\alpha_0 \,  
a_*^2}} \, \sinh{\left(H\tau \right)}\right]} \,.
\ee
The scale factor as a function of $\tau$ is then 
\begin{equation} 
a(\tau) = \sqrt{\frac{\beta_0}{\alpha_0}} \, 
\sinh{\left(H\tau \right)} \,.
\end{equation}

\subsection{Range of validity of the geodesic coordinates} 

Let us establish the range of validity of the Gautreau-Hoffmann-like  
geodesic coordinate patch. First, remember that the components of the 
four-velocity of radial geodesic observers are 
\begin{align}
u^t = & \ \sqrt{\frac{\beta_0 }{\alpha_0 \, a^2 \left(1 - k 
R^2/a^2 \right)} + 1} \,,\\ 
&\nonumber\\
u^R =& \ HR \sqrt{\frac{\beta_0 }{\alpha_0 \, a^2 \left(1 - k R^2/a^2 
\right)} + 1} \, - \frac{1}{a} \, \sqrt{\frac{\beta_0 }{\alpha_0 }} \,,
\end{align}
with $\alpha_0 $ and $ \beta_0 $  given by Eqs.~(\ref{alpha0}) and 
(\ref{beta0}). For a flat FLRW universe 
it is $1 - H_0^2 \, R_0^2 > 0$  and $a_0 \, H_0 \, R_0 \geq 0$ for $R_0 < 
1/H_0$, then the time component of the four-velocity is 
\be
u^t = \sqrt{ \frac{\beta_0}{ \alpha_0 \, a^2 }\, + 1} \,,
\ee
which is defined only for $ \beta_0 / \left(\alpha_0 \, a^2\right) > 
-1$. The radial component  $u^R$ of the four-velocity of a radial geodesic 
observer is defined only if  $\alpha_0 > 0$, and one concludes that the 
geodesic coordinates must satisfy $ R_0 <  1/H_0$.

For a curved ($k = \pm 1$) FLRW universe, $\alpha_0$ must be 
positive  again, which is equivalent to the constraint 
\be
1 - R_0^2  \left( H_0^2 + \frac{k}{a_0^2} \right) > 0 
\ee
or
\begin{equation}
    R_0 < \frac{1}{\sqrt{H_0^2 + k/a_0^2}} \,,
\end{equation}
where the right hand side is the radius of the apparent cosmological horizon 
\cite{Faraoni:2015ula}. If the universe is negatively curved, then 
\be
\beta_0 = H_0^2 
R_0^2  \left(a_0^2 + R_0^2 \right) > 0 
\ee
while, if it is positively curved, 
\be
\beta_0 = H_0^2 R_0^2 \left(a_0^2 - R_0^2 \right) \geq 0
\ee
for $R_0 \leq a_0$, which is always satisfied due to the 
constraint on $u^t$. The latter gives
\begin{equation}
    \frac{\beta_0 }{\alpha_0 \, a^2 \left(1 - k R^2/a^2 \right)} 
> 0\,.
\end{equation}

\subsection{Line element in geodesic coordinates}

Let us attempt to express the FLRW line element in Gautreau-Hoffmann-like  
geodesic coordinates.  Eq.~(\ref{relation:ttau}) can be used again to 
express $dt$ in terms of  $d\tau$, 
\begin{equation}
dt = u^t d\tau = \sqrt{\frac{\beta_0 }{\alpha_0 \, a^2 \left(1 - k 
R^2/a^2 \right)} + 1} \, \, d\tau 
\end{equation}
which, substituted in the FLRW line element in 
pseudo- Painlev\'e-Gullstrand coordinates~(\ref{pseudoPG}) 
produces the non-vanishing metric components 
\begin{align}
g_{\tau \tau} =& \ -\left(1-\frac{H^2R^2}{1-kR^2/a^2}\right) 
\left(\frac{\beta_0 }{\alpha_0 \, a^2 \left(1 - k R^2/a^2 \right)} + 
1 \right) \,,\\
&\nonumber\\
g_{\tau R} = & \ -\frac{HR}{1-kR^2/a^2} 
\sqrt{\frac{\beta_0 }{\alpha_0 \, a^2 \left(1 - k R^2/a^2 \right)} + 
1} \,,\\
&\nonumber\\
g_{RR} = & \ \frac{1}{1-kR^2/a^2} \,,\\
&\nonumber\\
g_{\vartheta \vartheta} =& \ R^2 \,,\\
&\nonumber\\
g_{\varphi \varphi} =& \ R^2 \sin^2{\vartheta} \,,
\end{align}
in geodesic coordinates, where now $a=a(t(\tau))$ and $H=H(t(\tau))$. When 
$d\tau = 0$, the Riemannian 3-spaces are the same as in the comoving FLRW 
foliation.

Let us consider again the special case of the de Sitter universe. Its line 
element in geodesic coordinates can be diagonalized by introducing a new 
radial coordinate $\rho = \rho(\tau,R)$ by
\begin{equation}
d\rho = \frac{1}{F}\left(\beta d\tau + dR \right) \,,
\end{equation}
where $\beta( \tau,R)$ must be determined {\em a posteriori} so that 
the cross-term $d\tau \, d\rho$ disappears, while $F(\tau, R)$ is an 
integrating factor satisfying
\begin{equation}
\frac{\partial}{\partial R} \left(\frac{1}{F} \right) = 
\frac{\partial}{\partial \tau} \left(\frac{\beta}{F} \right) 
\label{eqforF}
\end{equation}
in order to guarantee that $d\tau$ is an exact differential. Using $dR = F 
d\rho - 
\beta d\tau$, one obtains
\begin{align}
    ds^2 =& \ -\left[\left(1 - H^2 R^2 
\right)\coth^2{\left(H\tau\right)} \right. \\
&\nonumber\\
& \ \left. - \beta^2 - 2HR \coth{\left(H\tau\right)} \beta \right] d\tau^2\\
&\nonumber\\
& \ -2F\left[HR \coth{\left(H\tau\right)} + \beta  \right] d\rho d\tau + 
F^2 d\rho^2 + R^2 d\Omega_{(2)}^2 \,.
\end{align}
The choice
\be
\beta = -HR\coth{\left(H\tau\right)} \label{betachoice}
\ee
eliminates the time-radius cross-term and diagonalizes the line element, that 
becomes  
\begin{align}
    ds^2 = -\coth^2{\left(H\tau\right)} d\tau^2 + F^2 d\rho^2 + 
R^2(\tau, \rho) d\Omega_{(2)}^2 \,.
\end{align}
With the choice~(\ref{betachoice}) of $\beta$, the general solution of 
Eq.~(\ref{eqforF}) is  
\begin{equation}
 F(\tau,R) = \frac{A \exp{\left[-(\lambda/2) 
R^2\right]}}{\tanh{\left(H \tau \right)}} \left[\cosh{\left(H \tau  
\right)} \right]^{\lambda/H^2} \,,
\end{equation}
where $A$ is an integration constant and $\lambda$ is a separation constant 
(see~\ref{sec:appendix}). Setting $A = 1$ and $\lambda = 0$ so that 
\begin{equation}
F(\tau,R) = \coth{\left(H \tau \right)} 
\end{equation}
produces the diagonal de Sitter line element 
\begin{align}
ds^2 = \coth^2{\left(H\tau\right)} \left( d\rho^2 - d\tau^2 \right) + 
R^2(\tau, \rho) d\Omega_{(2)}^2 \,.
\end{align}

\section{Concluding remarks}
\label{sec:4}

We have revisited geodesic and quasi-geodesic observers in FLRW universes, 
removing the restrictions intrinsic in Gautreau's previous work, which was 
limited to de Sitter and Einstein-de Sitter universes 
\cite{Gautreau83,Gautreau84}. In general, geodesic coordinates turn out to be 
rather cumbersome in generic FLRW spaces, expecially those with curved spatial 
sections. In particular, one would like to express the geodesic time $\tau$ as 
a function of the comoving time $t$, the parameters, and the initial 
conditions along the radial timelike geodesics of FLRW space. The time 
measured by freely falling clocks ({\em i.e.}, the proper time of radial 
geodesic massive observers) is expressed by an integral that, in general, 
cannot be computed explicitly in terms of elementary functions, even in 
spatially flat FLRW universes. This situation, however, improves in most 
situations of practical interest, including the case of a power-law scale 
factor and, of course, in de Sitter space. We have provided explicit solutions 
for a radiation fluid, a stiff fluid, and empty de Sitter space. The latter, 
being locally static, is rather similar to the Schwarschild geometry and was 
already discussed by Gautreau \cite{Gautreau83}, who had already used 
geodesic coordinates in Schwarzschild space in his earlier joint paper with 
Hoffmann 
\cite{Gautreau:1978zz}. For power-law scale factors $a(t)$ in $k=0$ FLRW 
universes, we have identified all the situations in which the geodesic time 
$\tau$ can be expressed explicitly in terms of comoving time by making use of 
the Chebysev theorem of integration \cite{Chebysev,MarchisottoZakeri}, under 
the mild assumption that the equation of state parameter $w$ is a rational 
number. Alternatively, one can use the representation of the 
integral~(\ref{Cintegral}) in terms of a hypergeometric function and note that 
the assumptions of the Chebysev theorem leading to integrability are 
equivalent to the conditions for the truncation of the hypergeometric series 
to a finite sum (this mathematical condition was noted several times in the 
different context of two-fluid cosmologies 
\cite{Jacobs1968,McIntosh1972,McIntoshFoyster1972,Chen:2014fqa,Faraoni:2021opj}).

The range of validity of geodesic coordinates is also limited: the radial 
coordinate is restricted to the region below the apparent horizon of the FLRW 
universe and, therefore, it is not expected that geodesic coordinates will be 
useful for the thermodynamics of this apparent horizon since they cannot 
penetrate it (contrary to the Kruskal-Szekeres coordinates 
\cite{Kruskal,Szekeres}, the Painlev\'e-Gullstrand coordinates 
\cite{Painleve,Gullstrand}, or their Martel-Poisson generalization 
\cite{Martel:2000rn} in the Schwarzschild geometry). In this region below 
the apparent horizon, the geodesic coordinates describe the internal clock of 
dark matter or of free-falling test particles.

\begin{acknowledgements} 

This work is supported by the Natural Sciences \& Engineering Research Council 
of Canada (grant no. 2016-03803 to V.F.) and by Bishop's University.

\end{acknowledgements}


\appendix
\section{Integrating factor for the de Sitter universe}
\label{sec:appendix}

Using the variable  $u(\tau,R) \equiv 1/F$, we have 
\begin{equation}
\frac{\partial u}{\partial R} = \frac{\partial}{\partial \tau} 
\left[ -H \, R \, u \coth{\left(H\tau\right)} \right] \,;
\end{equation}
assume the ansatz
\be
u \left( \tau, R \right) = T(\tau) \, S(R) \,,
\ee
then it is
\be
T \, \frac{d S}{d R} =  - R \, S \, \frac{d}{d \tau} 
\left[ H\coth{\left(H\tau\right)} \, T \right] \,.
\ee
Dividing both sides by $R \, S \, T$ gives
\begin{equation}
 \frac{1}{R S} \, \frac{d S}{d R} = - \frac{1}{T} \, \frac{d}{d \tau} 
\left[ H\coth{\left(H\tau\right)} \, T \right] \,;
\end{equation}
the left-hand side depends only on $R$ while the right-side depends only 
on $\tau$, hence it must be 
\begin{equation}
\frac{1}{R S} \, \frac{d S}{d R} = \lambda = - \frac{1}{T} \, \frac{d}{d \tau} 
\left[H\coth{\left(H\tau\right)} \, T \right] \,,
\end{equation}
where $\lambda$ is a separation constant. The function $S(R)$ obeys 
\be
    \frac{1}{S} \, \frac{d S}{d R} =  \lambda R \,,
\ee
which integrates to 
\be
\ln{|S|} =  \frac{\lambda}{2} \, R^2 + C_1
\ee
(with $C_1$ an integration constant) and
\be
 S =  A_1 \exp{\left( \frac{ \lambda R^2}{2} \right) } \,.
\ee
The time part $T(\tau)$ satisfies the equation
\be
\frac{d}{d \tau} \left[ H\coth{\left(H\tau\right)} \, T \right] =  
- \lambda T \,,
\ee
which yields
\be
H \coth{\left(H\tau\right)} \, \frac{dT}{d \tau} = -\left(\lambda + 
\frac{d}{d \tau} \left[ H\coth{\left(H\tau\right)}\right]\right) T 
\ee
and 
\be     
\int{\frac{dT}{T}} = \ -\int{\frac{\lambda 
d\tau}{H\coth{\left(H\tau\right)}}} - 
\int{\frac{d\left( H\coth{\left(H\tau\right)} 
\right)}{H\coth{\left(H\tau\right)}}} \,,
\ee
giving
\be
T = \ \frac{A_2 }{H} \, \tanh{\left(H \tau \right)} 
\left[\cosh{\left( H \tau  \right)} \right]^{-\lambda/H^2} \,.
\ee
The general solution for $u(\tau, R)$ is, therefore, 
\begin{eqnarray}
u(\tau,R) &=& \frac{A_1 A_2}{H} \, \exp{\left( \frac{ \lambda  R^2}{2} 
\right) }  \tanh{\left(H \tau \right)} \left[\cosh{\left(H \tau 
\right)} \right]^{-\lambda/H^2} \,.\nonumber\\
&&
\end{eqnarray}
As $A_{1,2}$ and $H$ are constants, $F(\tau,R)$ is of the form
\begin{equation}
F(\tau,R) = \frac{A \, \exp{\left[-(\lambda/2) R^2\right]}}{ \tanh{\left(H 
\tau \right)}} \, \left[\cosh{\left(H \tau \right)} \right]^{\lambda/H^2} 
\,,
\end{equation}
where $A$ is an integration constant.



\end{document}